\documentclass[preprint,showkeys,amsmath,amssymb]{revtex4}

\usepackage[dvipdfm]{graphicx}
\usepackage{dcolumn}
\usepackage{bm}

\begin{document}

\title{Coexistence of Antiferromagnetism and Superconductivity in PrFeAsO$_{1-\delta}$}

\author{Daichi Kubota}
 \email{kubotada@pe.osakafu-u.ac.jp}
\author{Takekazu Ishida}
 \email{ishida@center.osakafu-u.ac.jp}
 \affiliation{%
Department of Physics and Electronics, and Institute for Nanofabrication Research, 
Osaka Prefecture University, 1-1, Gakuen-cho, Naka-ku, Sakai, Osaka 599-8531, Japan
}%
\author{Motoyuki Ishikado}
\author{Shin-ichi Shamoto}
\affiliation{%
Neutron Science Research Center, Japan Atomic Energy Research Institute, 2-4, Shirataka-Shirane, Tokai-mura, Naka-gun, Ibaraki 319-1195, Japan
}%

\author{Hiroshi Eisaki}
\author{Hijiri Kito}
\author{Akira Iyo}

\affiliation{%
Nanoelectronics Research Institute, National Institute of Advanced Industrial Science and Technology, 1-1-1, Umezono, Tsukuba, Ibaraki 305-8568, Japan
}%

\date{\today}

\begin{abstract}

A high-quality PrFeAsO$_{1-\delta}$ single crystal ($T_{\rm c}=44$ K) has been investigated by the magnetic torque.
Antiferromagnetism of the Pr$^{3+}$ ions was found to coexist with superconductivity in PrFeAsO$_{1-\delta}$ at temperatures below $T_\mathrm{N}=14$\,K.
We predict a magnetic structure that is not in accordance with earlier neutron studies performed  using polycrystalline nonsuperconducting specimens.
As the temperature decreases, the superconducting anisotropy $\gamma\sim 4$ of PrFeAsO$_{1-\delta}$ increases near $T_c$ and tends to decrease slightly at lower temperatures.

\end{abstract}

\pacs{74.25.Ha, 74.70.Ad}

\keywords{PrFeAsO$_{1-\delta}$, single crystal, torque, anisotropy, antiferromagnetism, weak ferromagnetism}

\maketitle

Since the discovery of superconductivity in the iron oxypnictide LaFeAsO$_{1-\delta}$F$_\delta$ \cite{HOSONOJACS2008} considerable effort has been made to understand the superconducting mechanism of this and similar materials.
Other iron oxypnictides have subsequently been reported to show $T_c$ as high as 55\,K and several promising applications in various fields using these materials have been identified.
The superconductivity in iron-oxypnictides has a rich variety of derivatives, i.e., the so-called 1111 phase \cite{TAKAHASHINATURE2008,CHENNATURE2008}, the 122 phase \cite{ROTTERPRL2008}, the 111 phase \cite{WANGSSC2008}, and the 11 phase \cite{HSUPNAS2008,TAKANOAPL2008}.

Achieving a high superconducting $T_c$ is a goal of contemporary scientific research.
An interesting finding is that antiferromagnetism is often closely tied with possible high-$T_c$.
For instance, the high-$T_c$ cuprate HgBa$_2$Ca$_4$Cu$_5$O$_y$ has five CuO$_2$ planes in a unit cell and its anisotropy ranges from 40 to 50 depending on the doping level \cite{CRISANPRB2007}.
$^{63,65}$Cu-NMR measurements of HgBa$_2$Ca$_4$Cu$_5$O$_y$ have revealed that disparate electronic phases emerge at the outer two superconducting CuO$_2$ planes and the inner three antiferromagnetic layers; the outer plane undergoes a bulk superconducting transition at $T_c=108$\,K and the underdoped inner plane shows a antiferromagnetic transition below $T_\mathrm{N}=60$\,K \cite{KOTEGAWAPRB2004}.

It is of great importance to determine whether magnetism and superconductivity can coexist in iron arsenic superconductors.
For example, a transition from a paramagnetic state to an antiferromagnetic state has been reported for both Fe$^{2+}$ and Pr$^{3+}$ ions. However, this is only the case for undoped non-superconducting samples \cite{ZHAOPRB08}.
Superconducting samples do not show antiferromagnetism.
McGuire {\it et al.} \cite{MCGUIRENJP09} studied M\"ossbauer spectroscopy of RE1111 samples (RE = La, Ce, Pr, Nd) and did not find any evidence for antiferromagnetism in La1111 and Nd1111, while they found $T_{\rm N}=3.8$ K for Ce1111 and $T_{\rm N} =13$ K for PrFeAsO$_{1-\delta}$.
Note that their studies were carried out using undoped specimens.

Zhao {\it et al.} \cite{ZHAONATUREMATERIALS2008} have reported systematic studies of CeFeAsO$_{1-x}$F$_x$ as a function of doping $x$, and confirmed three independent phase transitions, i.e., structural phase transition, antiferromagnetic transition of Fe spins, and antiferromagnetic transition of Ce spins.
To the authors' knowledge, the magnetic ordering in RE1111 systems has been reported only for undoped nonsuperconducting systems, by neutron diffraction \cite{ZHAOPRB08,ZHAONATUREMATERIALS2008,QIUPRL2008}, M\"ossbauer spectroscopy \cite{MCGUIRENJP09}, and $\mu$SR spectroscopy \cite{AMATOPHYSICAC2009,MAETERPRB2009}.
Neutron studies of magnetic ordering in RE1111 systems suggest that Fe$^{2+}$ spins are in an antiferromagnetic state along the $a$ axis while Pr$^{3+}$ spins are in an antiferromagnetic state along the $c$ axis \cite{ZHAOPRB08}.
Superconductivity only appears in a regime where antiferromagnetism disappears as a function of $x$.
These findings strongly suggest that it is very hard for iron arsenic superconductors to accommodate antiferromagnetism and superconductivity simultaneously.

In contrast to this result, recent $^{149}$Sm nuclear resonant forward scattering (NRFS) measurements by Kamihara {\it et al.} \cite{KAMIHARA_ARXIV2009} indicate that an antiferromagnetic Sm sublattice appears in the superconducting phase of SmFeAsO$_{1-x}$F$_x$  ($x = 0.069$) below $T_\mathrm{N} = 4.4$\,K while the undoped sample has $T_\mathrm{N} = 5.6$\,K.
Magnetism from Sm$^{3+}$ ions has also been reported in the optimal superconducting phase in SmFeAsO$_{1-x}$F$_x$ ($x = 0.15$) \cite{DINGPRB2009}.
Drew {\it et al.} \cite{DREWNATUREMATERIALS2009} have reported antiferromagnetic ordering of Sm spins below 5\,K by means of ZF-$\mu$SR measurements as well as specific heat measurements for SmFeAsO$_{1-x}$F$_x$ ($x = 0 - 0.2$).
However, no magnetic orderings have been observed so far for superconducting PrFeAsO$_{1-x}$F$_x$ \cite{ROTUNDUPRB2009}.
We argue that it is necessary to carry out further studies on the possible coexistence of superconductivity and antiferromagnetism using high-quality crystals of iron arsenic superconductors.

The magnetic torque is a good measure of electronic anisotropy, including magnetism and superconductivity.
In this Letter, we report clear evidence for the coexistence of antiferromagnetism and superconductivity from systematic measurements of the magnetic torque of a high-quality superconducting PrFeAsO$_{1-\delta}$ single crystal.

\begin{figure}[!ht]
\includegraphics[width=0.80\linewidth]{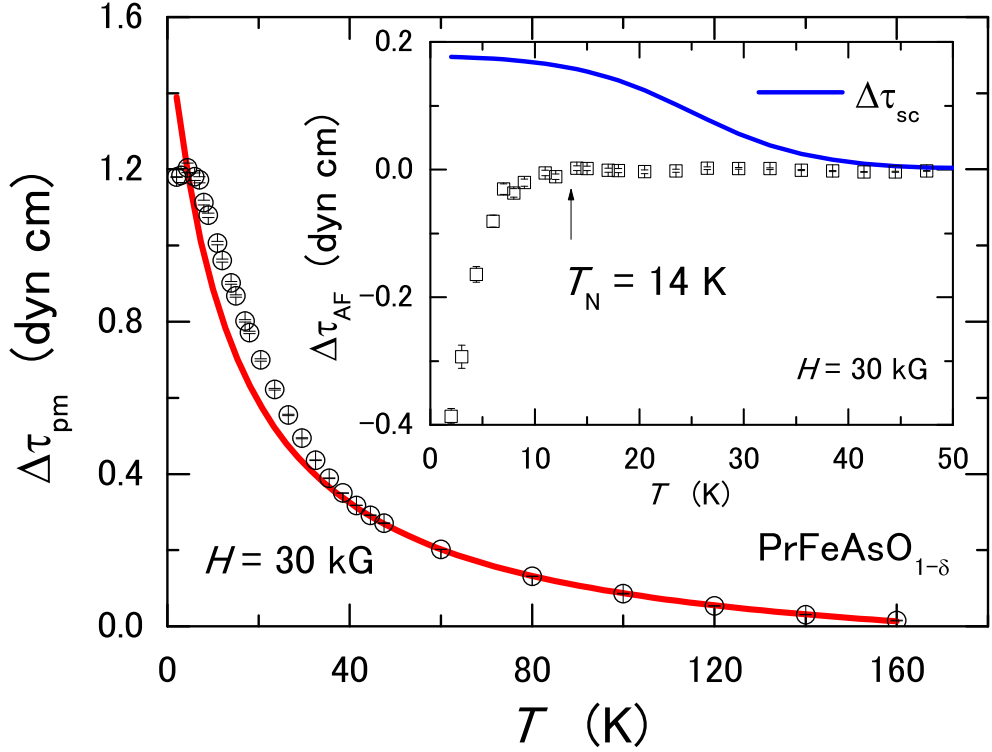}
\caption{
(Color online)
The $\sin 2\theta$ component of the PrFeAsO$_{1-\delta}$ torque as a function of $T$ in a magnetic field of 30\,kG. 
The solid line represents the least-squares fit to the Curie--Weiss law $\tau_{\rm 2\theta} = C/(T-\Theta)+\mathrm{\tau_0}$ ($C=24.4\pm 2.1$, $\Theta=14.1\pm 3.8$ K, $\tau_0=-0.13\pm0.01$) in the normal state.
After subtracting the superconducting torque $\Delta \tau_{\rm sc}$ of the inset (see text), the torque followed the curve $\Delta \tau_{\rm AF}$, demonstrating clear evidence for antiferromagnetism at temperatures below 14\,K.
}
\label{FIG1}
\end{figure}

The PrFeAsO$_{1-\delta}$ single crystal had a plate-like shape (560$\times$400\,$\mu$m in size, 19\,$\mu$m in thickness), of which the critical temperature  $T_c$ was determined to be 44 K by a SQUID magnetometer.
The torque is a bulk probe and is defined as the angular derivative of the free energy $F$ with respect to $\theta$, $\tau(\theta) = - \partial F/\partial \theta$.
The reversible torque is evaluated as $\tau_{\rm rev}(\theta_c) = (\tau_{\rm inc}(\theta_c)+\tau_{\rm dec}(\theta_c))/2$, where $\tau_{\rm inc}(\theta_c)$ and $\tau_{\rm dec}(\theta_c)$ represent the torque as functions of increasing and decreasing $\theta_{c}$, respectively.
$\theta_{c}$ is the angle between the applied magnetic field $H$ and the $c$-axis.
The anisotropic paramagnetism gives a torque expressed by 
$
\tau_{\rm pm}(\theta_{c})=\frac{1}{2}(\chi_a-\chi_c)H^2V\sin 2\theta_{c}=\Delta \tau_{\rm pm} \sin 2\theta_c
$
where $\chi_a$ and $\chi_c$ are the susceptibility along the $a$-axis and that along the $c$-axis, respectively.
The paramagnetic anisotropy in PrFeAsO$_{1-\delta}$ is presumably due to Pr$^{3+}$ spins.
In Fig.~\ref{FIG1}, the coefficient $\Delta \tau_{\rm pm}$ of $\sin 2\theta_c$ is plotted as a function of $T$ by analyzing the PrFeAsO$_{1-\delta}$ torque curves in a magnetic field of 30\,kG. The data is fitted to the Curie--Weiss law of Pr$^{3+}$ spins in the normal state from 47.5 to 180 K.

The $\sin 2\theta$ component consists of the anisotropic paramagnetism $\Delta \tau_{\rm pm}$, the {\it effective} excess contribution $\Delta \tau_{\rm sc}$ due to superconductivity, and the antiferromagnetism $\Delta \tau_{\rm AF}$ (see the inset of Fig.~\ref{FIG1}).
We find that the enhanced coefficient of $\sin\,2\theta$, in excess of that predicted by the Curie--Weiss law, is due to anisotropic superconductivity in PrFeAsO$_{1-\delta}$.
For simplicity, the superconducting torque was approximated by a presumed line $\Delta \tau_{\rm sc}=\Delta \tau_{\rm sc}^0 [ 1- \tanh (T-T_{\rm mid})/T_{\rm width}]$ ($\Delta\tau_{\rm sc}^0 = 0.09\pm 0.002$ dyncm, $T_{\rm mid}=25.1\pm 0.44$ K, $T_{\rm width}=11.3\pm0.5$ K), saturated at low temperatures.
Antiferromagnetism appears at temperatures below 14\,K after subtracting the superconducting torque $\Delta \tau_{\rm sc}$.
Our torque measurements show that the susceptibility follows the relation $\chi_a > \chi_c$ even at temperatures below $T_{\rm N}$.
This strongly indicates that the $c$ axis should be considered the easy axis of the antiferromagnetic state in PrFeAsO$_{1-\delta}$.
We argue that the spin configuration in the antiferromagnetic state is inconsistent with the findings of neutron diffraction experiments on a non-superconducting polycrystalline sample \cite{ZHAOPRB08}.
The sign of $\Delta \tau_{\rm 2\theta}$ in the torque curve remains unchanged in both the normal and superconducting states. 
Therefore, the direction of the antiferromagnetic spins should be parallel to the $ab$ plane. 
In view of the fact that the antiferromagnetic torque $\Delta\tau_{\rm AF}$ is appreciable compared to the superconducting torque $\Delta \tau_{\rm sc}$ (see inset of Fig.~\ref{FIG1}), the antiferromagnetism at temperatures below $T_c$ cannot be explained by an impurity phase.

\begin{figure}[!ht]
\includegraphics[width=0.8\linewidth]{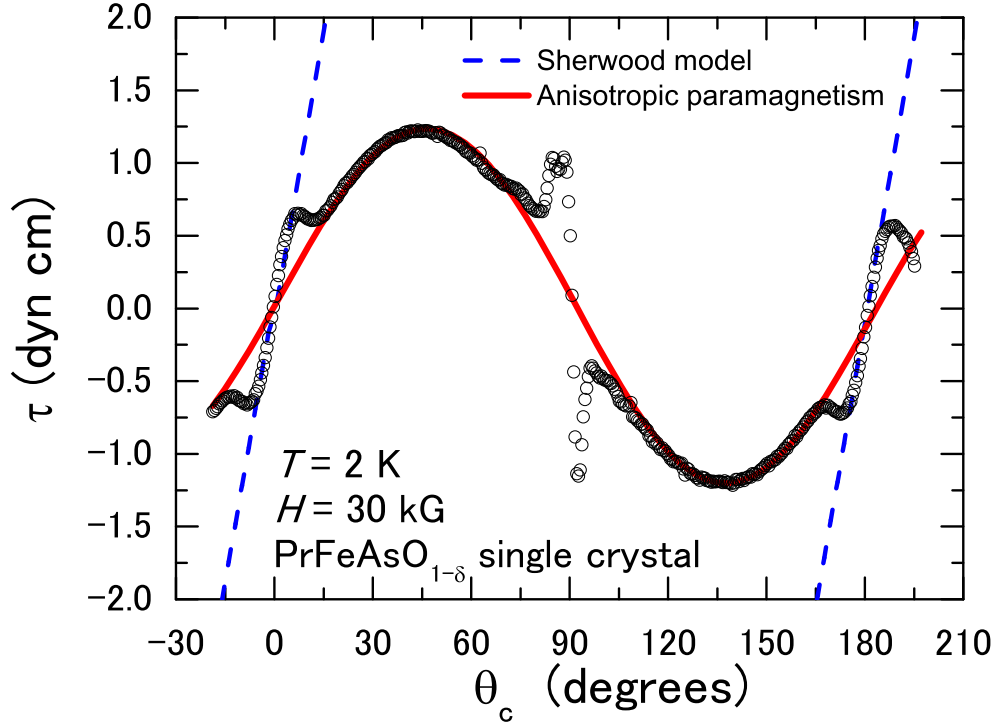}
\caption{
(Color online)
The reversible torque curve (open circles) in 30 kG and at 2 K.
The solid line shows a fit to $\tau (\theta_c) = \Delta \tau {\sin }2\theta_c$ ($\Delta \tau= 1.18\pm 0.01$).
The dashed line shows a fit to the Sherwood model $\tau_{\rm wf}(\theta_{c})=\pm \sigma_0HV\sin\theta_c$ ($\sigma_0=7.6\pm 1.9$) of excess torque coming from field-induced weak ferromagnetism.
}
\label{FIG2}
\end{figure}

Fig.~\ref{FIG2} shows the reversible torque of PrFeAsO$_{1-\delta}$ in 30 kG and at 2 K.
In addition to the $\sin 2\theta$ component, we observe that the torque exhibits an unstable nature at $\theta_c=0$. 
This unstable torque region appears only at temperatures lower than $T_{\rm N}=14$ K, and its magnitude increases as $T$ decreases.
We argue that this unstable torque is due to weak ferromagnetism of PrFeAsO$_{1-\delta}$ induced in the antiferromagnetic ordered state when the field is applied parallel to the $c$ axis.
Sherwood {\it et al.} \cite{SHERWOODJAP1959} proposed a torque model for field-induced weak ferromagnetism or parasitic magnetic ordering REFeO$_3$.
The torque formula is given by $\tau_{\rm wf}(\theta_{c})=\pm \sigma_{\theta}HV\sin\theta_c=\Delta \tau_{\theta} \sin \theta$, where the coefficient $\sigma_0$ is supposed to be independent of $H$ and the sign ($\pm$) depends on the field direction in the crystal.
Fig.~\ref{FIG2} also shows a fitting curve of the enhanced unstable torque to $\tau_{\rm wf}(\theta_{c})=\Delta \tau_{\theta} \sin \theta$.
Note that the formula becomes $\tau=\tau_{\rm 2\theta}\sin 2\theta$ for anisotropic paramagnetism.
The magnetic contribution of Tm$^{3+}$ in TmFeO$_3$ shows a similar anomalous torque curve when weak ferromagnetism is induced in the antiferromagnetic phase \cite{KURODAPR1961}.
Field-induced canted ferromagnetism is a candidate to explain weak ferromagnetism yielding the $\sin \theta$ term in PrFeAsO$_{1-\delta}$ expected from the Sherwood formula \cite{SHERWOODJAP1959}.
Both anisotropic paramagnetism and antiferromagnetism yield the $\sin 2\theta_c$ term, arising from the difference in the susceptibilities along the $ab$-plane and the $c$-axis.
It is worth noting that the $\sin\,2\theta$ term is proportional to $H^2$ while the $\sin \theta$ term due to spontaneous magnetization is proportional to $H^1$.
We compared the prefactor $\tau_0$ in 30\,kG with that in 10\,kG to investigate the origin of the anomalous magnetism.
The ratio $\Delta\tau_{\rm 2\theta}(H=30\, \mathrm{kG})/\Delta \tau_{2\theta}(H=10\, \mathrm{kG})$ remains at $2.1\pm 0.3$ while it is expected to reach 9.
The ratio $\Delta\tau_{\rm \theta}(H=30\, \mathrm{kG})/\Delta \tau_{\theta}(H=10\, \mathrm{kG})$ is $2.7\pm 0.5$, in good accordance with the theoretical expectation of 3.
We reasonably conclude that the origin of the singular torque curve for PrFeAsO$_{1-\delta}$ at $\theta_c$ = 0 degrees is due to field-induced canted ferromagnetism.

Heat capacity measurements on LaFeAsO$_{1-x}$F$_x$ by Kohama {\it et al.} \cite{KOHAMAPRB2008} suggest the occurrence of itinerant ferromagnetism.
The appearance of ferromagnetic spin fluctuations in LaFeAsO$_{1-x}$F$_x$ is interpreted as being due to electronic interactions between itinerant quasiparticles.  
This anomaly indicates that ferromagnetic interaction between itinerant quasiparticles in the superconducting state in the Fe--As layers gives rise to a weak ferromagnetic state when an external field is applied to the system.
A similar weak ferromagnetic spin fluctuation has been reported to occur in superconducting UCoGe \cite{HUYPRL2007}.
The fact that ferromagnetic instability occurs in the vicinity of the superconducting state strongly suggests the occurrence of triplet superconductivity in UCoGe.
It is not certain whether a similar scenario occurs in PrFeAsO$_{1-\delta}$, and further studies are needed.


\begin{figure}[!ht] 
\includegraphics[width=0.8\linewidth]{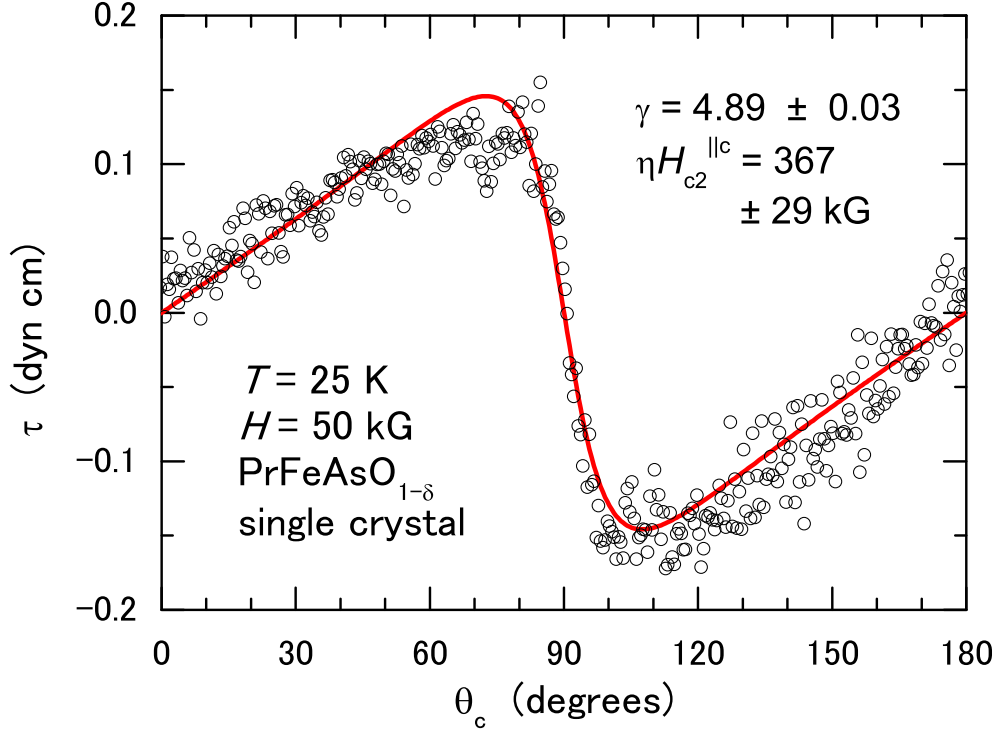}
\caption{
(Color online)
The reversible torque curve of the PrFeAsO$_{1-\delta}$ single crystal in 50\,kG and at 25\,K, where the paramagnetic contribution to the torque is subtracted.
The solid line is the least-square fit to the Kogan model.
}
\label{FIG3}
\end{figure}

The PrFeAsO$_{1-\delta}$ single crystal does not exhibit conventional intrinsic pinning above 14 K, while some high-$T_{\rm c}$ cuprates show an extremely sharp peak in the torque hysteresis at $\theta_c=90$ degrees \cite{ISHIDAPRB1997B,ISHIDAPRB1998,NARIAKIPHYSICAC2001,NARIAKIPHYSICAC2002}.
This is due to undulation of the order parameter perpendicular to the superconducting planes.
The absence of conventional intrinsic pinning in the torque curve indicates modest anisotropy in oxypnictides.
However, a novel intrinsic pinning behavior can be seen at temperatures below 14 K.
As seen in Fig.~\ref{FIG2}, a steep stable point can be seen at $\theta_c$ = 90 degrees of the torque curve.
The torque hysteresis also shows a sharp peak at 90 degrees.
This is very similar to the so-called intrinsic pinning that appears in various high-$T_c$ cuprates \cite{ISHIDAPRB1997B,ISHIDAPRB1998,NARIAKIPHYSICAC2001,NARIAKIPHYSICAC2002}. 
We consider that this is due to the magnetic interaction between vortices and antiferromagnetic spins.
When the field is exactly parallel to the antiferromagnetic plane, the induced moment along the field may decrease the free energy compared to the paramagnetic case.
This is also a sort of exotic intrinsic pinning, as first discovered by us at temperatures below $T_{\rm N}$.

Conventional torque theory for anisotropic superconductors, developed by Kogan \cite{KOGANPRB1998}, is very useful for various types of superconductors \cite{FARRELLPRL1990,ISHIDAPRB1997B,ISHIDAPRB1998,NARIAKIPHYSICAC2001,NARIAKIPHYSICAC2002}. It is given by 
\begin{eqnarray}
\tau_{\rm rev}(\theta_{c})&=&
\frac{\Phi_0 H V}{16\pi\lambda^2}
\frac{\gamma^2-1}{\gamma^{1/3}}
\frac{\sin 2\theta_{c}}{\epsilon(\theta_{c})} 
\ln \{{{\gamma \eta H_{c2}^{\parallel c}} 
\over {H\epsilon(\theta_{c}})}\} 
\label{Kogan eq}
\end{eqnarray}
where $\epsilon(\theta_{c}) = (\sin^2\theta_{c}+\gamma^2 \cos^2\theta_{c})^{1/2}$, $\gamma=\sqrt{m_c/m_{ab}}$, $\Phi_0$ is a flux quantum, $H_{c2}^{\parallel c}$ is the upper critical field when the field is applied parallel to the c axis ($\eta$\,$\sim$\,1), and $V$ is the sample volume.

\begin{figure}
\includegraphics[width=0.8\linewidth]{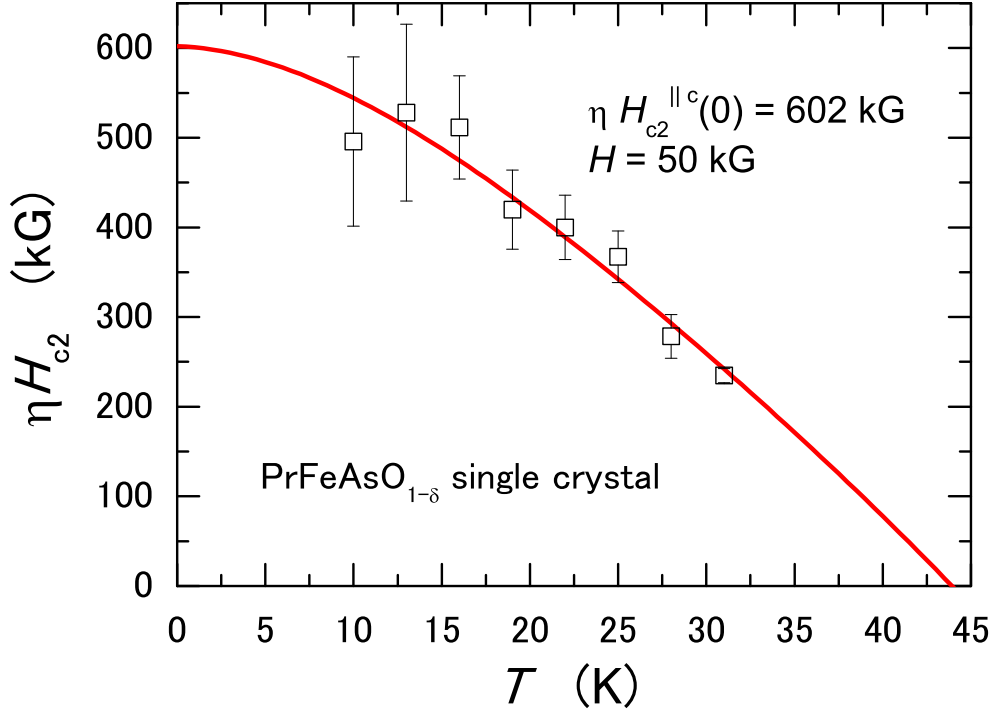}
\caption{
(Color online)
$\eta H_{c2}$ in 50 kG determined by the Kogan model as a function of $T$.
The solid line is the least-square fit to WHH theory ( $\alpha = 0$, $\lambda_\mathrm{SO} = 0$) \cite{WHH_PR1966}.
}
\label{FIG4}
\end{figure}

Fig.~\ref{FIG3} shows a typical reversible superconducting torque curve of the PrFeAsO$_{1-\delta}$ crystal at 25\,K and in 50\,kG.
The paramagnetic contribution to the PrFeAsO$_{1-\delta}$ torque curve was estimated in the regime between  $T_c$ and 180 K, and the superconducting torque was extracted from the raw torque curve.
The correction for antiferromagnetism is also taken into account. 
In the analysis of the torque curves in 50\,kG, both $\gamma$ and $\eta H_{c2}^{\parallel c}$ are treated as free parameters in the least-square fitting.
In Fig.~\ref{FIG4}, $\eta H_{c2}^{\parallel c}$ as a function of temperature in 50\,kG is approximated by a least-square fit to Werthamer--Helfand--Hohenberg (WHH) theory under the conditions $\alpha = 0$ and $\lambda_\mathrm{SO} = 0$ \cite{WHH_PR1966}.
We predict $\eta H_{c2}^{\parallel c}(0) = 602$\,kG at $T=0$\,K from the fitted curve.
In Fig.~\ref{FIG4}, the solid line is a least-square fit curve by means of the single-band Kogan model with fixed $\eta H_{c2}$ of Eq.~(\ref{Kogan eq}), where $\gamma$ is the only fitting parameter and $\eta H_{c2}^{\parallel c}$ is taken from the solid line of Fig.~\ref{FIG4}.

\begin{figure}
\includegraphics[width=0.8\linewidth]{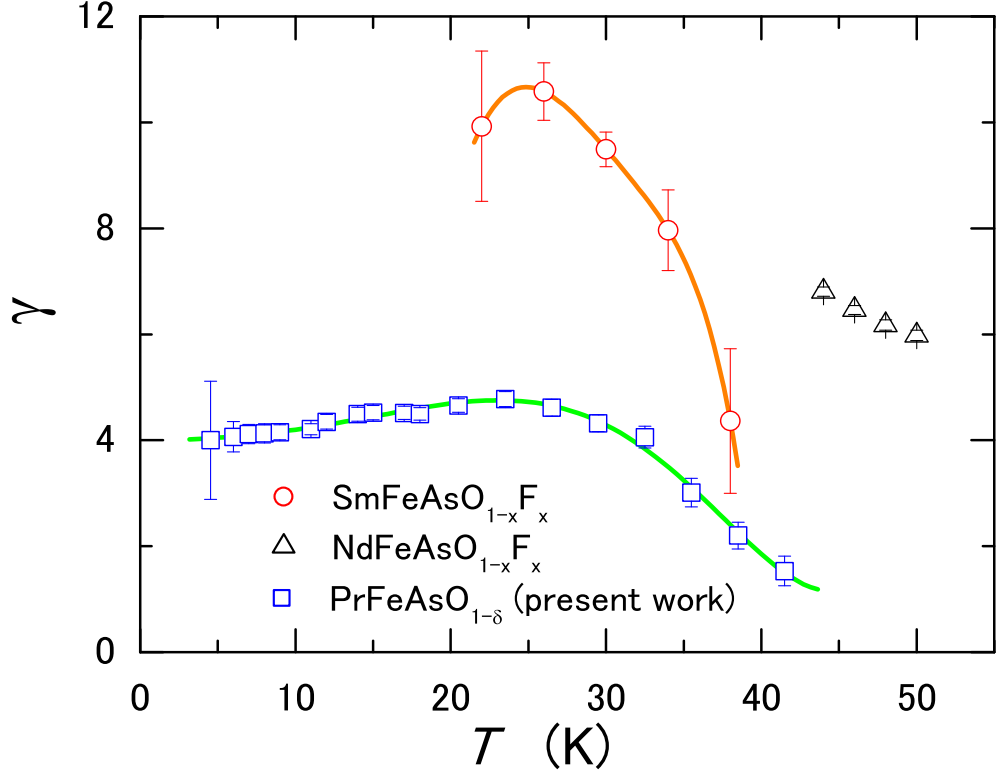}
\caption{(Color online) 
The anisotropy $\gamma$ for PrFeAsO$_{1-\delta}$ determined in the present work (open squares). 
Data taken from SmFeAsO$_{1-x}$F$_{x}$ torque analysis in 30\,kG (open circles) are also shown for comparison \cite{BALICAS_ARXIV2008}. 
The anisotropy determined from the angular dependence of the resistivity of a NdFeAsO$_{1-x}$F$_{x}$ single crystal from 5\,kG to 90\,kG is also shown (open triangles) \cite{JIASSUST2009}.
The lines are guides for the eye.
}
\label{FIG5}
\end{figure}

We analyzed the torque using the Kogan model.
In Fig.~\ref{FIG5}, we show the $\gamma$ value of the PrFeAsO$_{1-\delta}$ single crystal compared with an undoped SmFeAsO$_{1-x}$F$_{x}$ single crystal ($T_\mathrm{c} =45$\,K)  \cite{BALICAS_ARXIV2008} as a function of temperature in 30\,kG
The $\gamma$ value of PrFeAsO$_{1-\delta}$ increases slightly from 4 to 5 as $T$ increases to 25\,K while the $\gamma$ of SmFeAsO$_{1-x}$F$_{x}$ varies from 4 to 11.
The anisotropy of NdFeAsO$_{1-x}$F$_{x}$ single crystals ($T_\mathrm{c} = 51.5$\,K) is also reported to increase from 4 to 6 ($5 < H < 90$\,kG), determined from the angular dependence of the resistivity \cite{JIASSUST2009},
$\gamma$ of PrFeAsO$_{1-\delta}$ increases gradually from 4 to 4.7 between 4 K and 25 K, and then decreases down to 1.5 between 25 K and 41.5 K.
SmFeAsO$_{1-x}$F$_{x}$ also exhibits a decrease in $\gamma$ as $T$ increases near $T_c$.
A consideration of the effect of multiband superconductivity in PrFeAsO$_{1-\delta}$ \cite{OKAZAKIPRB2009} is of interest, and the theory for a multi-band superconductor has been reported \cite{KOGANPRL2002,KOGANPRB2004} and an improved theory will be published elsewhere \cite{KUBOTA_TOBEPUBLISHED2010}.

In conclusion, the anisotropy parameter $\gamma$ for PrFeAsO$_{1-\delta}$ is modest and changes gradually with temperature.
After correcting the effect of the paramagnetic contribution on the torque curve, we can analyze the superconducting anisotropy $\gamma$ using the Kogan model.
A very interesting finding is that magnetic ordering at $T_{\rm N}$ = 14 K appears even in the superconducting PrFeAsO$_{1-\delta}$ sample.
This provides crucial evidence that superconductivity and antiferromagnetism can coexist at temperatures below $T_\mathrm{N}$ even in iron arsenic superconductors.
Moreover, the appearance of field-induced weak ferromagnetism at temperatures below $T_{\rm N}$ further confirms the appearance of antiferromagnetism.

This work was supported in part by a Grant-in-Aid for Scientific Research from the Ministry of Education, Culture, Sports, Science and Technology of Japan (Grant No. 19206104) and a special grant from Osaka Prefecture University.





\begin{thebibliography}{9}

\bibitem{HOSONOJACS2008}
Y. Kamihara {\it et al.}, 
J. Am. Chem. Soc. \textbf{130}, 3296 (2008).

%
\bibitem{TAKAHASHINATURE2008}
H. Takahashi {\it et al.}, 
Nature {\bf 453}, 376 (2008).

%
\bibitem{CHENNATURE2008}
X. H. Chen {\it et al.}, 
Nature {\bf 453}, 761 (2008).

%
\bibitem{ROTTERPRL2008}
M. Rotter, M. Tegel, and D. Johrendt, 
Phys. Rev. Lett. {\bf 101}, 107006 (2008).

%
\bibitem{WANGSSC2008}
X. C. Wang {\it et al.}, 
Solid State Comm. {\bf 148}, 538 (2008).

%
\bibitem{HSUPNAS2008}
F. -C. Hsu {\it et al.}, 
Proc. Natl. Acad. Sci. {\bf 105}, 14262 (2008).

%
\bibitem{TAKANOAPL2008}
Y. Mizuguchi {\it et al.}, 
Appl. Phys. Lett. {\bf 93}, 152505 (2008).


%
\bibitem{CRISANPRB2007}
A. Crisan {\it et al.}, 
Phys. Rev. B {\bf 76}, 212508 (2007).

%
\bibitem{KOTEGAWAPRB2004}
H. Kotegawa {\it et al.}, 
Phys. Rev. B {\bf 69}, 014501 (2004).

%
\bibitem{ZHAOPRB08}
J. Zhao {\it et al.}, 
Phys. Rev. B {\bf 78}, 132504 (2008).

%
\bibitem{MCGUIRENJP09}
M. A. McGuire {\it et al.}, 
New J. Phys. {\bf 11}, 025011 (2009).

%
\bibitem{ZHAONATUREMATERIALS2008}
J. Zhao {\it et al.}, 
Nature Materials {\bf 7}, 953 (2008).

%
\bibitem{QIUPRL2008}
Y. Qiu {\it et al.}, 
Phys. Rev. Lett. {\bf 101}, 257002 (2008).

%
\bibitem{AMATOPHYSICAC2009}
A. Amato {\it et al.}, 
Physica C {\bf 469}, 606 (2009).

%
\bibitem{MAETERPRB2009}
H. Maeter {\it et al.}, 
Phys. Rev. B {\bf 80}, 094524 (2009).

%
\bibitem{KAMIHARA_ARXIV2009}
Y. Kamihara {\it et al.}, 
arXiv:0904.3173v2 (2009).

%
\bibitem{DINGPRB2009}
L. Ding {\it et al.}, 
Phys. Rev. B {\bf 77}, 180510(R) (2008).

%
\bibitem{DREWNATUREMATERIALS2009}
A. J. Drew {\it et al.}, 
Nature Materials {\bf 8}, 310 (2009).

\bibitem{ROTUNDUPRB2009}
C. R. Rotundu {\it et al.}, 
Phys. Rev. B {\bf 80}, 144517 (2009).


%
\bibitem{SHERWOODJAP1959}
R. C. Sherwood {\it et al.}, 
J. Appl. Phys. {\bf 30}, 217 (1959).

%
\bibitem{KURODAPR1961}
C. Kuroda {\it et al.}, 
Phys. Rev. {\bf 122}, 446 (1961).

%
\bibitem{KOHAMAPRB2008}
Y. Kohama {\it et al.}, 
Phys. Rev. B {\bf 78}, 020512(R) (2008).

%
\bibitem{HUYPRL2007}
N. T. Huy {\it et al.}, 
Phys. Rev. Lett. {\bf 99}, 067006 (2007).

\bibitem{ISHIDAPRB1997B}
T. Ishida {\it et al.}, 
Phys. Rev. B {\bf 56}, 11897 (1997).

%
\bibitem{ISHIDAPRB1998}
T. Ishida, K. Okuda, A. I. Rykov, S. Tajima, and I. Terasaki, 
Phys. Rev. B. {\bf 58}, 5222 (1998).


%
\bibitem{NARIAKIPHYSICAC2001}
N. Yamamoto {\it et al.}, 
Physica C {\bf 357-360}, 298 (2001).

%
\bibitem{NARIAKIPHYSICAC2002}
N. Yamamoto {\it et al.}, 
Physica C {\bf 378-381}, 483 (2002).


%
\bibitem{KOGANPRB1998}
V. G. Kogan, Phys. Rev. B {\bf 38}, 7049 (1988). 

%
\bibitem{FARRELLPRL1990}
D. E. Farrell, J. P. Rice, D. M. Ginsberg, and J. Z. Liu, 
Phys. Rev. Lett. {\bf 64}, 1573 (1990).

%
\bibitem{WHH_PR1966}
N. R. Werthamer, E. Helfand, and P. C. Hohenberg, 
Phys. Rev. {\bf 147}, 295 (1966).

%
\bibitem{BALICAS_ARXIV2008}
L. Balicas {\it et al.}, 
arXiv:0809.4223v2 (2008).

%
\bibitem{JIASSUST2009}
Y. Jia {\it et al.}, 
Supercond. Sci. Technol. {\bf 21}, 105018 (2008).

%
\bibitem{OKAZAKIPRB2009}
R. Okazaki {\it et al.}, 
Phys. Rev. B {\bf 79}, 064520 (2009).

%
\bibitem{KOGANPRL2002}
V. G. Kogan, Phys. Rev. Lett. {\bf 89}, 237005 (2002).

%
\bibitem{KOGANPRB2004}
V. G. Kogan and N. V. Zhelezina, Phys. Rev. B {\bf 69}, 132506 (2004).

%
\bibitem{KUBOTA_TOBEPUBLISHED2010}
D. Kubota, N. Hayashi, and T. Ishida, to be published.

\end{thebibliography}
\end{document}